\documentclass[a4paper]{cas-dc}
\usepackage[numbers,sort&compress]{natbib}
\usepackage{braket}
\usepackage{multicol}
\usepackage{subfigure}
\usepackage{ulem}

\definecolor{darkpastelgreen}{rgb}{0.01, 0.75, 0.24}
\definecolor{electricindigo}{rgb}{0.44, 0.0, 1.0}
\definecolor{palatinateblue}{rgb}{0.15, 0.23, 0.89}
\definecolor{carminered}{rgb}{1.0, 0.0, 0.22}
\hypersetup{linkcolor=palatinateblue, citecolor=electricindigo, urlcolor=carminered}



\begin{document}
\shorttitle{Scale-anomaly-induced binding pressure in hadrons}
\shortauthors{D. Fujii and M. Tanaka}  
\title[mode = title]{Scale-anomaly-induced binding pressure in hadrons}

\author[1,2]{Daisuke Fujii}[type=editor, orcid=0000-0002-6298-9278]
\nonumnote{$\dag$ E-mail: daisuke@rcnp.osaka-u.ac.jp (corresponding author)}

\author[3]{Mitsuru Tanaka}[orcid=0009-0000-5341-562X]
\nonumnote{$\ddag$ E-mail: tanaka@hken.phys.nagoya-u.ac.jp (corresponding author)}

\address[1]{Advanced Science Research Center, Japan Atomic Energy Agency (JAEA), Tokai, 319-1195, Japan}
\address[2]{Research Center for Nuclear Physics, Osaka University, Ibaraki 567-0048, Japan}
\address[3]{Department of Physics, Nagoya University, Nagoya 464-8602, Japan}

\begin{abstract}
The effect of the QCD scale anomaly on the internal pressure distribution of hadrons is studied based on the trace-traceless decomposition of the energy-momentum tensor. Using recent model-independent results of gravitational form factors as input, the pressure distributions of both pions and nucleons are analyzed in the instant form and the light-front form. It is found that, in all cases, the scale anomaly dominantly generates the negative binding pressure. This result suggests that the phenomenon is a universal feature, independent of models, types of hadrons, and the choice of form.
\end{abstract}


\maketitle

\section{Introduction}

The scale anomaly in Quantum Chromodynamics (QCD) plays a crucial role in ensuring the existence of hadrons, and it is measured by the trace of the energy-momentum tensor (EMT). The forward limit of the hadronic matrix element of the EMT is related to the mass of the hadron, and it is known that the scale anomaly generates a significant contribution to the hadron mass. 
A deeper understanding of the role played by the scale anomaly thus provides a vital key to unveiling the origin of hadron structure.

The hadronic matrix elements of the EMT contain not only the information on mass but also rich details of internal structure. The associated gravitational form factors (GFFs) encode the spatial distributions of energy density, spin density, and stress, which consists of pressure and shear forces. In particular, the stress distribution has attracted considerable attention from both theoretical~\cite{Polyakov:1999gs,Brommel:2005ee,Brommel:2007zz,LHPC:2007blg,Broniowski:2008hx,Frederico:2009fk,Masjuan:2012sk,Son:2014sna,Yang:2014xsa,Bali:2016wqg,Fanelli:2016aqc,Hudson:2017xug,Shanahan:2018pib,Polyakov:2018exb,Shanahan:2018nnv,Lorce:2018egm,Alexandrou:2018xnp,Alexandrou:2019ali,Anikin:2019kwi,Avelino:2019esh,Azizi:2019ytx,Freese:2019bhb,Freese:2019eww,Hatta:2019lxo,Mamo:2019mka,Neubelt:2019sou,Yanagihara:2019foh,Alharazin:2020yjv,Varma:2020crx,Chakrabarti:2020kdc,Kim:2020lrs,Kim:2020nug,Krutov:2020ewr,Shuryak:2020ktq,Yanagihara:2020tvs,Alexandrou:2020sml,deTeramond:2021lxc,Freese:2021czn,Freese:2021qtb,Gegelia:2021wnj,Hatta:2021can,Ji:2021mfb,Kim:2021jjf,Loffler:2021afv,Lorce:2021xku,Mamo:2021krl,More:2021stk,Owa:2021hnj,Panteleeva:2021iip,Pefkou:2021fni,Raya:2021zrz,Tong:2021ctu,Alharazin:2022wjj,Choudhary:2022den,Fujita:2022jus,Kim:2022wkc,Lorce:2022cle,Mamo:2022eui,Tanaka:2022wzy,Won:2022cyy,Ji:2022exr,Fu:2022rkn,Amor-Quiroz:2023rke,Czarnecki:2023yqd,Guo:2023pqw,Guo:2023qgu,Hackett:2023nkr,Hackett:2023rif,Hatta:2023fqc,Ito:2023oby,Liu:2023cse,Lorce:2023zzg,Li:2023izn,Xu:2023izo,Won:2023ial,Won:2023zmf,Fu:2023ijy,Wang:2023bjp,Broniowski:2024oyk,Cao:2024zlf,HadStruc:2024rix,Liu:2024jno,Liu:2024vkj,Yao:2024ixu,Fujii:2024rqd,Freese:2024rkr,Broniowski:2025ctl,Cao:2025dkv,Dehghan:2025eov,Dehghan:2025ncw,Ghim:2025gqo,Goharipour:2025lep,Goharipour:2025yxm,Guo:2025jiz,Hatta:2025ryj,Hatta:2025vhs,Liu:2025vfe,Nair:2025sfr,Sugimoto:2025btn,Ji:2025gsq,Fujii:2025aip,Fujii:2025tpk,Tanaka:2025pny} and experimental~\cite{Belle:2012wwz,Belle:2015oin,Kumano:2017lhr,Burkert:2018nvj,AbdulKhalek:2021gbh,Anderle:2021wcy,Duran:2022xag,Krutov:2022zgg,JeffersonLabHallA:2022pnx,CLAS:2022syx} perspectives since it was first extracted from experimental data for the proton in 2018~\cite{Burkert:2018bqq} (for reviews, see Refs.~\cite{Polyakov:2018zvc,Burkert:2023wzr}).

The contribution of the scale anomaly to the EMT was first elucidated in a pioneering study~\cite{Ji:1994av}, which proposed a four-term decomposition of the hadron mass based on the trace-traceless decomposition of the EMT; this framework was later extended to the spin decomposition~\cite{Ji:1996ek}. We note that, whereas the trace of the EMT, given by the sum of the gluonic scale anomaly and the quark-mass term accompanied by an anomalous dimension, is itself independent of renormalization scheme and scale, the individual parts in the four-term decomposition have been criticized for their scheme and scale dependence and for operator mixing under renormalization, thereby motivating alternative mass-decomposition prescriptions~\cite{Metz:2020vxd,Lorce:2021xku} (see also Ref.~\cite{Ji:2021mtz} for a response to these criticisms). In view of the above, in this letter we do not enter into any particular decomposition prescription; instead, we clarify the role of the scale anomaly by examining the effect of the EMT trace on the stress distributions. 

Previous theoretical studies, including our own work~\cite{Fujii:2025aip,Fujii:2025tpk,Tanaka:2025pny} as well as those by other groups~\cite{Ji:2025gsq}, have investigated the impact of the scale anomaly on the pressure distribution. These studies have demonstrated that the contribution from the scale anomaly plays a dominant role in generating the negative pressure--namely, the binding pressure, following Ref.~\cite{Burkert:2018bqq}. In all of these analyses, the pressure distribution were evaluated in the instant form, through which the contribution of the scale anomaly to the pressure has been identified. 
The instant-form framework allows for a three-dimensional visualization of the internal stress distribution of hadrons; however, its definition depends on the choice of equal-time hypersurfaces and the inertial frame, which imposes limitations on describing the internal structure in a Lorentz-invariant manner~\cite{Miller:2018ybm,Jaffe:2020ebz}. 

Therefore, in this study, we analyze the role of the scale anomaly in the stress distributions of nucleons and pions, not only in the instant form but also in the light-front (LF) form. In contrast, the LF form enables a Lorentz-invariant definition of two-dimensional densities on the transverse plane at fixed LF time, providing a natural connection to the impact-parameter representation of generalized parton distributions (GPDs), which are experimentally accessible~\cite{Burkardt:2000za,Burkardt:2002hr,Diehl:2002he,Miller:2018ybm,Jaffe:2020ebz,Freese:2021czn,Freese:2021mzg,Freese:2021qtb,Cao:2023ohj,Freese:2023abr}.

In this study, we analyze the stress distributions of nucleons and pions in both the instant form and the LF form, using recently obtained model-independent results for GFFs~\cite{Cao:2024zlf}. This suggests the universality of the binding pressure driven by the scale anomaly, irrespective of types of hadrons (pions or nucleons) and the form (instant or LF). These findings shed new light on the universal and form-independent role of the scale anomaly in hadron structure, providing important guidance for future theoretical and experimental investigations.

\section{Definition of the stress tensor density on the instant and the LF form}

In this letter, we elucidate the role of the QCD scale anomaly in generating a stable pressure distribution inside hadrons. To this end, we decompose the EMT $T^{\mu\nu}(x)$ based on the trace-traceless separation as follows:
\begin{align}
   &T^{\mu\nu}=\bar{T}^{\mu\nu}+\hat{T}^{\mu\nu} \\
   &\bar{T}^{\mu\nu}=T^{\mu\nu}-\frac{1}{4}\eta^{\mu\nu}{T^\rho}_\rho \\
   &\hat{T}^{\mu\nu}=\frac{1}{4}\eta^{\mu\nu}{T^\rho}_\rho, 
\end{align}
where $\bar{T}^{\mu\nu}$ and $\hat{T}^{\mu\nu}$ denote the traceless and trace parts, respectively. In this work, we adopt the mostly-minus metric convention $\eta^{\mu\nu} = {\rm diag}(+1, -1, -1, -1)$. The momentum eigenstates are normalized as $\langle \vec{p}^{\prime} | \vec{p} \rangle = 2p^0 (2\pi)^3 \delta^{(3)}(\vec{p} - \vec{p}^{\prime})$, where $p^0 = \sqrt{\vec{p}^{,2} + m^2}$ and $m$ denotes the hadron mass, with $m = m_\pi$ for pions or $m = m_N$ for nucleons.

The EMT matrix elements of the pion are characterized by two independent form factors, $A^\pi(t)$ and $D^\pi(t)$, and are defined as
\begin{align}
    &\braket{\pi^a(p^\prime)|\bar{T}^{\mu\nu}(x)|\pi^b(p)} \notag \\
    &=\delta^{ab}\Big[2P^\mu P^\nu A^\pi(t)+\frac{1}{2}(\Delta^\mu\Delta^\nu-\eta^{\mu\nu}\Delta^2)D^\pi(t) \notag \\
    &\hspace{10mm}-2\frac{m_\pi^2}{4}G_s^\pi(t)\eta^{\mu\nu}\Big]e^{i(p^\prime-p)x} \\
    &\braket{\pi^a(p^\prime)|\hat{T}^{\mu\nu}(x)|\pi^b(p)}=\delta^{ab}\left[2\frac{m_\pi^2}{4}G_s^\pi(t)\eta^{\mu\nu}\right]e^{i(p^\prime-p)x},
\end{align}
where, the momenta $P^\mu$ and $\Delta^\mu$ are defined as $P^\mu = (p^\mu + {p^\prime}^\mu)/2$ and $\Delta^\mu = {p^\prime}^\mu - p^\mu$, with $t = -\Delta^2$. The scalar form factor $G_s^\pi(t)$, which characterizes the trace part of the EMT, is given in terms of $A^\pi(t)$ and $D^\pi(t)$ by 
\begin{align}
    G_s^\pi(t) = A^\pi(t) - \frac{\Delta^2}{4m_\pi^2}\left(A^\pi(t)+D^\pi(t)\right).
\end{align}
\begin{align}
    G_s^\pi(t) = A^\pi(t) - \frac{\Delta^2}{4m_\pi^2}\left(A^\pi(t)+3D^\pi(t)\right).
\end{align}

The EMT matrix elements of the nucleon are characterized by three independent form factors, $A^N(t)$, $J^N(t)$, and $D^N(t)$, and are given by
\begin{align}
    &\braket{N(p^\prime,s^\prime)|\bar{T}^{\mu\nu}(x)|N(p,s)} \notag \\
    &=\bar{u}^\prime\Big[A^N(t)\frac{P_\mu P_\nu}{m_N}+J^N(t)\frac{i\left(P_\mu\sigma_{\nu\rho}+P_\nu\sigma_{\mu\rho}\right)\Delta^\rho}{2m_N} \notag \\
    &\hspace{4mm}+D^N(t)\frac{\Delta_\mu\Delta_\nu-\eta_{\mu\nu}\Delta^2}{4m_N}-\frac{m_N}{4}G_s^N(t)\eta^{\mu\nu}\Big]ue^{i(p^\prime-p)x}, \label{GFFs} \\
    &\braket{N(p^\prime,s^\prime)|\hat{T}^{\mu\nu}(x)|N(p,s)}=\bar{u}^\prime\left[\frac{m_N}{4}G_s^N(t)\eta^{\mu\nu}\right]ue^{i(p^\prime-p)x},
\end{align}
where, $\sigma_{\mu\nu}$ is defined by $\sigma_{\mu\nu} = \frac{i}{2}[\gamma^\mu, \gamma^\nu]$, where $\gamma^\mu$ are the Dirac matrices, and $u(p,s)$ denotes the Dirac spinor with spin projections $s, s' = \pm 1/2$. We adopt the spinor normalization $\bar{u}(p)u(p) = 2p^0$.
The form factor $G_s^N(t)$, which characterizes the trace part of the EMT, is expressed for the nucleon as 
\begin{align}
    G_s^N(t)=A^N(t)-\frac{\Delta^2}{4m_N^2}\left(A^N(t)-2J^N(t)+3D^N(t)\right).
\end{align}

From Poincaré symmetry, GFFs $A^\pi(t)$, $A^N(t)$, and $J^N(t)$ satisfy the conditions $A^\pi(0) = A^N(0) = 1$, and $J^N(0) = 1/2$ in the forward limit. In contrast, there is no such constraint for the D-term at zero momentum transfer, $D^\pi(0)$ and $D^N(0)$. As discussed in Ref.~\cite{Perevalova:2016dln}, many systems are found to have a negative D-term. However, the sign of the D-term is not determined by mechanical stability in general. The hydrogen atom is a stable counterexample with $D(0)>0$~\cite{Ji:2022exr,Czarnecki:2023yqd,Freese:2024rkr}. For spin-$3/2$ baryons such as the $\Delta$ and $\Omega^-$, $D(0)>0$ is likewise reported~\cite{Fu:2022rkn,Fu:2023ijy,Wang:2023bjp}.

In conventional treatments, spatial densities associated with a given local operator have been defined as the three-dimensional Fourier transforms of form factors that characterize matrix elements between plane-wave states at fixed instant form time. However, it has been pointed out that such densities can only be interpreted as physically meaningful spatial distributions when the target particle is sufficiently heavy and a nonrelativistic description is valid~\cite{Miller:2018ybm,Jaffe:2020ebz}. In other cases, relativistic corrections become unavoidable. Moreover, since the choice of equal-time hypersurface is not invariant under Lorentz boosts in the instant form, contributions from different time slices cannot be eliminated, introducing an inherent ambiguity in the definition of these densities~\cite{Miller:2018ybm,Jaffe:2020ebz}.

In contrast, it is known that such ambiguities can be entirely avoided in the LF formulation with fixed LF time~\cite{Burkardt:2000za,Burkardt:2002hr,Diehl:2002he,Miller:2018ybm,Jaffe:2020ebz,Freese:2021czn,Freese:2021mzg,Freese:2021qtb,Cao:2023ohj,Freese:2023abr}. Therefore, following Ref.~\cite{Lorce:2018egm}, we define the spatial densities of pions and nucleons from both the instant form and LF form perspectives.

We begin by defining the three-dimensional density distribution in the instant form. At fixed instant form time ($x^0 = 0$), the spatial density of the stress tensor, $S^{ij}_{\rm BF}(\vec{x})$, can be written in the Breit frame $\vec{P} = 0$ (i.e., $\Delta_0 = 0$) as
\begin{align}
S^{ij}_{\rm BF}(\vec{x})=\int\frac{d^3\vec{\Delta}}{2P^0(2\pi)^3}e^{-i\vec{x}\cdot\vec{\Delta}}\braket{\vec{\Delta}/2|T^{ij}(0)|-\vec{\Delta}/2}, \label{staticEMT}
\end{align}
with $P^0 = \sqrt{m^2 + \vec{\Delta}^2/4}$. Here, we focus on the spatial components of the EMT, and only diagonal matrix elements without helicity flip contribute. So, we adopt the representative case with $s = s' = 1/2$. This expression is therefore applicable to both pions and nucleons. 
For spatially spherically symmetric systems such as the pion and nucleon, the pressure in the Breit frame is defined as $p_{\rm BF}(r) = \delta_{ij}S^{ij}_{\rm BF}/3 = \bar{p}_{\rm BF}(r) + \hat{p}_{\rm BF}(r)$, where $\bar{p}_{\rm BF}(r)$ and $\hat{p}_{\rm BF}(r)$ represent the pressure components derived from $\bar{T}^{\mu\nu}$ and $\hat{T}^{\mu\nu}$, respectively. The corresponding pressure distributions for the pion and nucleon are expressed as 
\begin{align}
    &\bar{p}_{\rm BF}^\pi(r)=\int\frac{d^3\vec{\Delta}}{2P^0(2\pi)^3}e^{-i\vec{\Delta}\cdot\vec{x}} \notag \\
    &\hspace{12mm}\times\left[\frac{m_\pi^2}{2}A^\pi+\frac{\vec{\Delta}^2}{24}(3A^\pi+D^\pi)\right] \label{barpBFpi} \\
    &\hat{p}_{\rm BF}^\pi(r)=\int\frac{d^3\vec{\Delta}}{2P^0(2\pi)^3}e^{-i\vec{\Delta}\cdot\vec{x}} \notag \\
    &\hspace{12mm}\times\left[-\frac{m_\pi^2}{2}A^\pi-\frac{\vec{\Delta}^2}{8}(A^\pi+3D^\pi)\right] \label{hatpBFpi} \\
    &\bar{p}_{\rm BF}^N(r)=\int\frac{d^3\vec{\Delta}}{2m_N(2\pi)^3}e^{-i\vec{\Delta}\cdot\vec{x}} \notag \\
    &\hspace{12mm}\times\left[\frac{m_N^2}{2}A^N+\frac{\vec{\Delta}^2}{24}(3A^N-6J^N+D^N)\right] \label{barpLFpi}\\
    &\hat{p}_{\rm BF}^N(r)=\int\frac{d^3\vec{\Delta}}{2m_N(2\pi)^3}e^{-i\vec{\Delta}\cdot\vec{x}} \notag \\
    &\hspace{12mm}\times\left[-\frac{m_N^2}{2}A^N-\frac{\vec{\Delta}^2}{8}(A^N-2J^N+3D^N)\right], \label{hatpLFpi}
\end{align}
where $r = |\vec{x}|$ is the radial distance.

Next, we define the two-dimensional transverse spatial density in the LF  form. In the LF form, the coordinates are defined as $x^\pm \equiv (x^0 \pm x^3)/\sqrt{2}$ and $\vec{x}_\perp \equiv (x^1, x^2)$. The longitudinal momentum is given by $P^+ \equiv \int dx^- d^2x_\perp T^{++} = (P^0 + P^3)/\sqrt{2}$, and $\vec{\Delta}_\perp = (\Delta^1, \Delta^2)$. 
At fixed LF time ($x^+ = 0$), by imposing $\Delta^+ = (\Delta^0 + \Delta^3)/\sqrt{2} = 0$ and integrating out $x^-$, the spatial density of the stress tensor, $S^{ab}_{\rm LF}(\vec{x}_\perp)$, is written as
\begin{align}
&S^{ab}_{\rm LF}(\vec{x}_\perp)= \notag \\
&\int\frac{d^2\vec{\Delta}_\perp}{2P^+(2\pi)^2}e^{-i\vec{x}_\perp\cdot\vec{\Delta}_\perp}\braket{P^+,\vec{\Delta}_\perp/2|T_{\rm LF}^{ab}(0)|P^+,-\vec{\Delta}_\perp/2}, \label{staticEMT}
\end{align}
where $T_{\rm LF}^{\alpha\beta}(x^+, \vec{x}_\perp) = \int dx^- T^{\alpha\beta}(x^+, x^-, \vec{x}_\perp)$, with $\alpha,\beta = +,1,2$ and $a,b = 1,2$. As in the instant form case, we take $s = s' = 1/2$, and thus this expression is also applicable to the pion. 
For spatially spherically symmetric systems such as the pion and nucleon, the pressure is defined as $p_{\rm LF}(x_\perp) = \delta_{ab}S^{ab}_{\rm LF}/2 = \bar{p}_{\rm LF}(x_\perp) + \hat{p}_{\rm LF}(x_\perp)$, and the pressure distributions for the pion and nucleon, are given by
\begin{align}
    &\bar{p}_{\rm LF}^\pi(x_\perp)=\int\frac{d^2\vec{\Delta}_\perp}{2P^+(2\pi)^2}e^{-i\vec{\Delta}_{\perp}\cdot\vec{x}_\perp} \notag \\
    &\hspace{12mm}\times\left[\frac{m_\pi^2}{2}A^\pi+\frac{\vec{\Delta}_\perp^2}{8}(A^\pi+D^\pi)\right] \label{barpBFN} \\
    &\hat{p}_{\rm LF}^\pi(x_\perp)=\int\frac{d^2\vec{\Delta}_\perp}{2P^+(2\pi)^2}e^{-i\vec{\Delta}_\perp\cdot\vec{x}_\perp} \notag \\
    &\hspace{12mm}\times\left[-\frac{m_\pi^2}{2}A^\pi-\frac{\vec{\Delta}_\perp^2}{8}(A^\pi+3D^\pi)\right] \label{hatpBFN} \\
    &\bar{p}_{\rm LF}^N(x_\perp)=\int\frac{d^2\vec{\Delta}_\perp}{2P^+(2\pi)^2}e^{-i\vec{\Delta}_{\perp}\cdot\vec{x}_\perp} \notag \\
    &\hspace{12mm}\times\left[\frac{m_N^2}{2}A^N\frac{\vec{\Delta}_\perp^2}{8}(A^N-2J^N+D^N)\right] \label{barpLFN} \\
    &\hat{p}_{\rm LF}^N(x_\perp)=\int\frac{d^2\vec{\Delta}_\perp}{2P^+(2\pi)^2}e^{-i\vec{\Delta}_{\perp}\cdot\vec{x}_\perp} \notag \\
    &\hspace{12mm}\times\left[-\frac{m_N^2}{2}A^N-\frac{\vec{\Delta}_\perp^2}{8}(A^N-2J^N+3D^N)\right], \label{hatLFFN}
\end{align}
where, $\bar{p}_{\rm LF}(x_\perp)$ and $\hat{p}_{\rm LF}(x_\perp)$ represent the pressure components derived from $\bar{T}^{\alpha\beta}$ and $\hat{T}^{\alpha\beta}$, respectively, and $x_\perp = |\vec{x}_\perp|$.

\section{Results: Contribution of the scale anomaly to the pressure}

In this study, we calculate the pressure distributions in the instant form and light-front (LF) form, as defined in Eqs.\eqref{barpBFpi}-\eqref{hatpLFpi} and \eqref{barpBFN}-\eqref{hatLFFN}, using as input the GFFs of the pion and nucleon that were determined in a model-independent manner in Ref.~\cite{Cao:2024zlf}, based on dispersion relations.\footnote{In this work, we focus on the qualitative behavior and do not take into account the uncertainty bands presented in Ref.~\cite{Cao:2024zlf}.}

To compute the pressure distributions, it is necessary to extrapolate the GFFs obtained in Ref.~\cite{Cao:2024zlf} to the asymptotic region of the momentum transfer $t$. In this study, following Refs.\cite{Lorce:2018egm,Freese:2021czn}, we employ the multipole model
\begin{align}
    F(t)=\frac{F(0)}{(1+t/\Lambda^2)^m}
\end{align}
for the extrapolation. This type of functions admits analytic expressions under Fourier transformation, making it suitable for qualitatively evaluating the features of the pressure distributions. With the modified Bessel function of the second kind $K_m$, for the three-dimensional Fourier transform, we have
\begin{align}
&\int\frac{d^3\vec\Delta}{(2\pi)^3}\,
\frac{e^{-i\vec\Delta\cdot\vec x}}{\bigl(1+\vec\Delta^{2}/\Lambda^2\bigr)^m} \notag \\
&\hspace{16mm}=
\frac{\Lambda^3}{4\pi^{3/2}}
\left(\frac{\Lambda\,r}{2}\right)^{m-3/2}\frac
{K_{\,m-3/2}\bigl(\Lambda\,r\bigr)}{(m-1)!},
\end{align}
and for the two-dimensional Fourier transform, we also have
\begin{align}
    \int\frac{d^2\vec{\Delta}_\perp}{(2\pi)^2}\frac{e^{-i\vec{\Delta}_\perp\cdot\vec{x}_\perp}}{(1+\vec{\Delta}^2_\perp/\Lambda^2)^m}=\frac{\Lambda}{\pi x_\perp}\left(\frac{\Lambda x_\perp}{2}\right)^m\frac{K_{m-1}(\Lambda x_\perp)}{(m-1)!}. \notag
\end{align}
In the present analysis, we adopt the tripole model with $m = 3$, since we have confirmed that the results shown below do not change qualitatively for $m = 2, 3, 4$.

As a result of fitting the GFFs provided in Ref.~\cite{Cao:2024zlf}, the parameter $\Lambda$ is determined as shown in Table~\ref{tab1}. Here, all the form factor values at $t = 0$ are fixed by known constraints, except for $D(0)$, which is treated as a free parameter\footnote{By approximate chiral symmetry, $D(0)$ is constrained to be close to $-1$ (see Refs.~\cite{Voloshin:1980zf,Novikov:1980fa}), with which our value, $D(0)=-0.935$, is consistent. In the exact chiral limit, one has $D(0)=-1$. 
Furthermore, our recent top-down holographic QCD analysis of the pion GFFs in the chiral limit likewise yields $D(0)=-1$~\cite{Fujii:2024rqd}.}. 
\begin{table}[t!]
\centering
\caption{Values of the parameter $\Lambda$ used in the multipole extrapolation of the GFFs. The fitting is performed using the model-independent GFFs from Ref.\cite{Cao:2024zlf}, which are based on dispersion relations.}
\begin{tabular}{lcccccc}
\hline\hline
& $m$ & Forward limit values: $F(0)$ & $\Lambda^2/{\rm GeV^2}$ \\
\hline
$A^\pi(t)$ & 3 & $1$ & 6.344 \\
$D^\pi(t)$ & 3 & $-0.935$ & 2.593 \\
$A^N(t)$ & 3 & $1$ & 3.213 \\
$J^N(t)$ & 3 & $1/2$ & 3.394 \\
$D^N(t)$ & 3 & $-3.352$ & 0.767 \\
\hline\hline
\end{tabular}
  \label{tab1}
\end{table}
With this setup, we are now ready to analyze the pressure distributions.

\begin{figure}
    \includegraphics[scale=0.3]{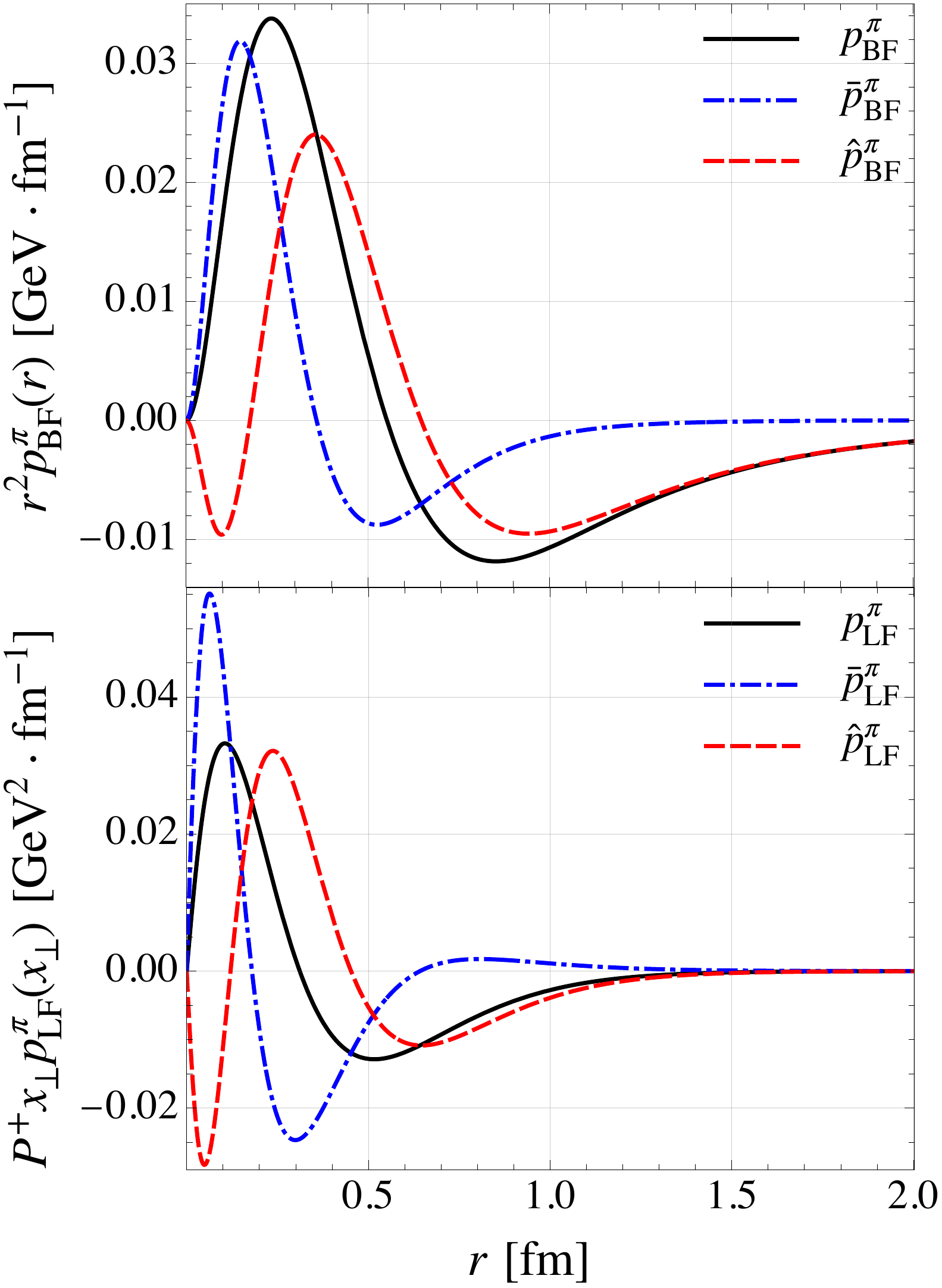}
    \caption{Pressure decomposition for pions.}
    \label{pressure_decomp_pion}
\end{figure}

\begin{figure}
    \includegraphics[scale=0.3]{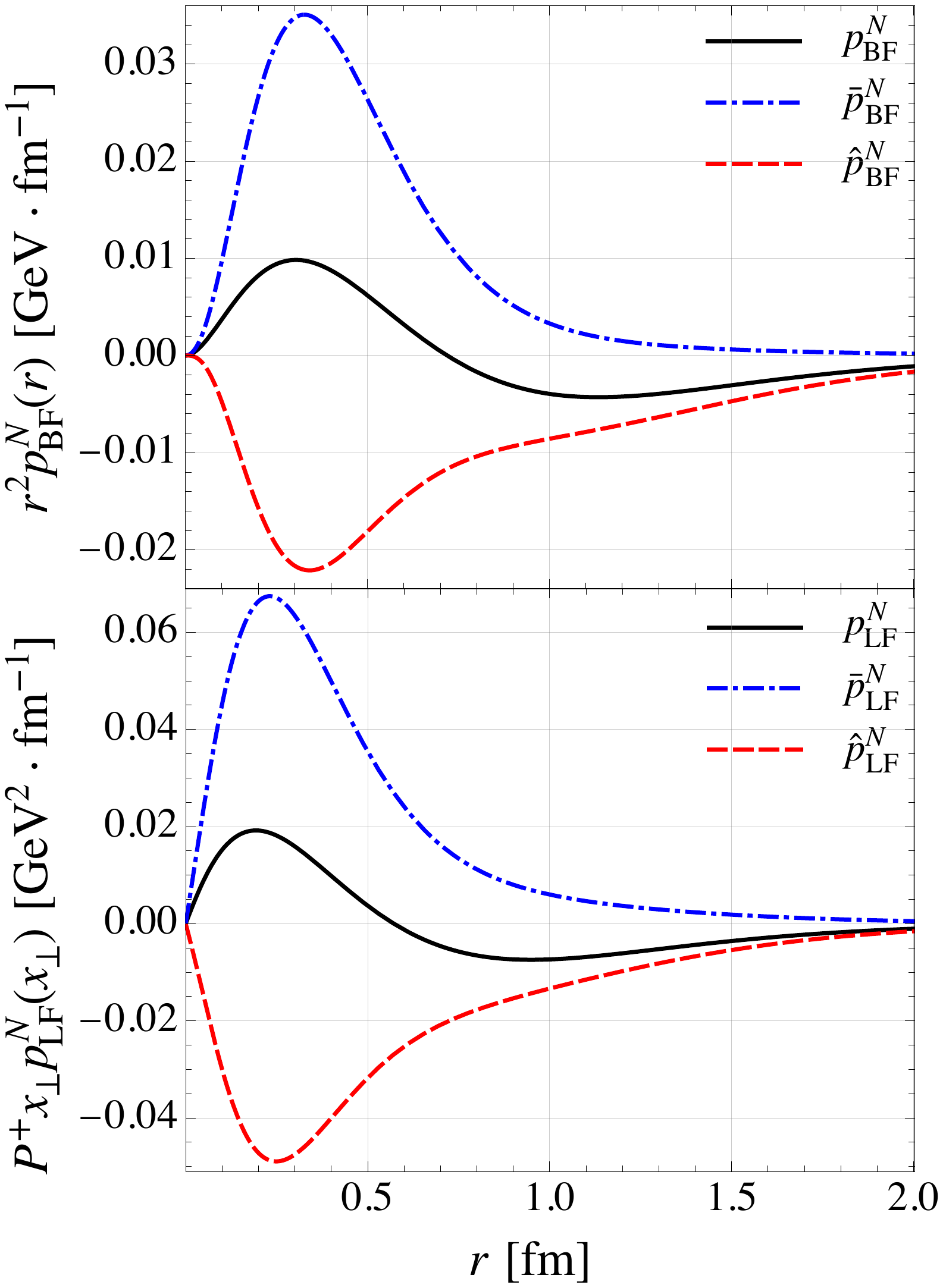}
    \caption{Pressure decomposition for nucleons.}
    \label{pressure_decomp_nucleon}
\end{figure}

First, we present the pressure distribution of the pion and its decomposition in the Breit frame on the instant form, shown in the upper panel of the Fig.~\ref{pressure_decomp_pion}. 
Here, the pion mass is fixed at the empirical value $m_\pi = 140 \ {\rm MeV}$. 
The total pressure distribution $p^\pi_{\rm BF}$ is positive near the center of the pion, exhibiting a repulsive behavior, while it becomes negative near the periphery, indicating an inward force that confines quarks and gluons inside the pion. Such a behavior of the pressure distribution is consistent with results observed in many previous studies.

Here, we focus on the decomposed pressure distributions. Near the center of the pion, $\bar{p}^\pi_{\rm BF}$ generates an outward pressure, while $\hat{p}^\pi_{\rm BF}$, originating from the scale anomaly\footnote{$\hat{p}^\pi$ includes not only the contribution from the scale anomaly but also the contribution from the current quark mass term. In this sense, it does not represent a purely the contribution from the scale anomaly. However, as confirmed in Ref.~\cite{Fujii:2025tpk}, we have verified that $\hat{p}^\pi$ exhibits similar behavior of the results in this study even in the chiral limit. Furthermore, for the nucleon, the contribution from the scale anomaly is dominant compared to that from the quark mass term. Therefore, in this work, the combined contribution from the quark mass term and the scale anomaly is referred to as the contribution from the scale anomaly.}, produces an inward binding pressure. In the intermediate region, we find that both $\bar{p}^\pi_{\rm BF}$ and $\hat{p}^\pi_{\rm BF}$ change their signs. As discussed in Ref.~\cite{Fujii:2025tpk}, this is a distinctive behavior that was not observed in previous studies~\cite{Fujii:2024rqd,Fujii:2025aip,Tanaka:2025pny} for the nucleon in the instant form, and it appears to be a unique feature of hadrons with a small mass, such as the pion. Near the pion surface, both $\bar{p}^\pi_{\rm BF}$ and $\hat{p}^\pi_{\rm BF}$ change sign again, with $\hat{p}^\pi_{\rm BF}$ once more driving the binding pressure. These results are similar to those obtained in Ref.~\cite{Fujii:2025tpk} for the massless pion in the LF form, suggesting that this behavior is a characteristic feature of light hadrons, independent of whether the instant or LF form is used.

However, as mentioned earlier, for light-mass particles such as the pion, the definition of spatial densities in the instant form still involves ambiguities. Therefore, performing a similar analysis in the LF form is essential for capturing the intrinsic features of the pion’s spatial structure.

The results for the pressure distribution of the pion and its decomposition in the LF form are shown in the lower part of the Fig.~\ref{pressure_decomp_pion}. While slight differences are observed in the absolute magnitude of the pressure and its spatial extent along the $r$ direction, the qualitative behavior closely resembles that found in the instant form. In particular, the behavior of $\hat{p}^\pi_{\rm LF}$, when considered together with the results in Ref.~\cite{Fujii:2025tpk}, strongly suggests that the pressure driven by the scale anomaly constitutes a binding force, and this feature is universal, independent of whether the analysis is performed in the instant or LF form.

Next, the pressure distribution of the nucleon in the Breit frame on the instant form and its decomposition are shown in the upper part of Fig.~\ref{pressure_decomp_nucleon}. Here, the nucleon mass is fixed to the empirical value $m_N = 940\ {\rm MeV}$. The total pressure distribution $p^N_{\rm BF}$ exhibits a repulsive behavior near the center, while it becomes negative near the surface, indicating an inward binding force. Such a behavior of the pressure distribution has been observed in many previous studies.

Focusing on the decomposed pressure distributions, $\bar{p}^N_{\rm BF}$ is found to take positive values for all radial distances, acting as a locally repulsive force, while the pressure component $\hat{p}^N_{\rm BF}$, originating from the scale anomaly, takes negative values and drives a binding force within the nucleon.
These features were first observed in our previous study based on a topological soliton model~\cite{Fujii:2024rqd}, and the present results strongly suggest that such behavior is a universal phenomenon, independent of the specific model employed. 

Finally, the pressure distribution of the nucleon and its decomposition in the LF form are shown in the lower panel of Fig.~\ref{pressure_decomp_nucleon}. While there are slight differences in the absolute magnitude and the radial extent of the pressure, the qualitative behavior is similar to the result obtained in the instant form. In particular, the fact that the pressure $\hat{p}^N_{\rm LF}$ induced by the scale anomaly also drives the binding pressure in the LF form supports results discovered in our previous study~\cite{Fujii:2025aip,Tanaka:2025pny}. 

\section{Summary and outlooks}

In this study, we computed the pressure distributions for the pion and nucleon in both the instant form and LF form, using as input the gravitational form factors determined in a model-independent manner based on  dispersion relations~\cite{Cao:2024zlf}. We investigated the role of the scale anomaly in these distributions. Our results show that in all cases, the pressure induced by the scale anomaly drives a negative binding pressure. A similar finding was previously observed in our earlier work~\cite{Fujii:2025aip,Fujii:2025tpk,Tanaka:2025pny}, and the present results strongly suggest that the phenomenon of scale anomaly-driven confinement pressure is not model-dependent. Moreover, the fact that this feature appears for both the pion and the nucleon indicates that it is a property independent of the type of hadron.
Furthermore, it is known that for particles such as the pion and nucleon, whose Compton wavelengths are comparable to their radii, the definition of densities in the instant form entails ambiguities~\cite{Miller:2018ybm,Jaffe:2020ebz}. Therefore, the fact that the same phenomenon is observed in both the instant form and the LF form in the present study strongly reinforces the claims made in our previous work~\cite{Fujii:2025aip,Tanaka:2025pny}. 

On the other hand, while the scale anomaly universally drives the binding pressure in both the pion and the nucleon, our analysis reveals notable differences in its detailed behavior between them. 
Therefore, as a future work, we will investigate the scale anomaly-driven confinement pressure of other hadrons and elucidate the origins of their characteristic features. 

\section*{Acknowledgments}

The authors thank Xiangdong Ji, Mamiya Kawaguchi, and Chen Yang for fruitful discussions about stress distributions. 
The author M.T. would like to take this opportunity to thank the financial support from "THERS Make New Standards Program for the Next Generation Researchers" and JST SPRING, Grant Number JPMJSP2125.
This work of D.F. was supported in part by the Japan Society for the Promotion of Science (JSPS) KAKENHI (Grants No. JP24K17054) and the COREnet project of RCNP, Osaka University.

  \setcounter{section}{0}
  \setcounter{equation}{0}
  \setcounter{figure}{0}
  \renewcommand{\theequation}{A\arabic{equation}}
  \renewcommand{\thesection}{A\arabic{section}}
  \renewcommand{\thefigure}{A\arabic{figure}}


\bibliographystyle{apsrev4-1}
\bibliography{ref}

  \setcounter{section}{0}
  \setcounter{equation}{0}
  \setcounter{figure}{0}
  \renewcommand{\theequation}{S\arabic{equation}}
  \renewcommand{\thesection}{S\arabic{section}}
  \renewcommand{\thefigure}{S\arabic{figure}}



\end{document}